\documentclass[aps,prd,twocolumn,superscriptaddress,floatfix,nofootinbib,showpacs,amsmath,amsfonts,amssymb,altaffilletter]{revtex4-2}
\pdfoutput=1
\UseRawInputEncoding
\usepackage{graphicx}
\usepackage{bm}
\usepackage{subfigure}
\usepackage{url}
\usepackage[hyperindex]{hyperref}
\usepackage{color}
\usepackage[ddmmyy,24hr]{datetime}
\usepackage{bigdelim}
\usepackage{booktabs}
\usepackage{dcolumn}
\usepackage{multirow}
\usepackage{subfigure}
\usepackage{physics}
\usepackage{cancel}
\usepackage{stackrel}
\usepackage{paralist}
\usepackage{xspace}
\usepackage{slashed}
\usepackage{cancel}
\usepackage{todonotes}
\usepackage{enumerate}
\usepackage{enumitem}
\usepackage{float}
\usepackage{fullpage}
\usepackage{ulem}
\usepackage{wasysym}
\usepackage{comment}
\usepackage{bbold}
\usepackage{orcidlink}
\usepackage{array}
\usepackage{color}
\usepackage[explicit]{titlesec}
\usepackage[utf8]{inputenc}
\usepackage{blindtext}
\usepackage{rotating}
\newcommand{\nua}[1]{\ensuremath{\rlap{\kern-2.5pt\ensuremath{\overset{\scriptscriptstyle(-)}{\phantom{\nu}}}}{\ensuremath{{\nu}_{#1}}}}}

\newcommand{\R}{R}
\begin{document}

\title{Revival of the Reactor Antineutrino Anomaly}

\author{Carlo Giunti}
\email{carlo.giunti@to.infn.it}
\affiliation{Istituto Nazionale di Fisica Nucleare (INFN), Sezione di Torino, Via P. Giuria 1, I--10125 Torino, Italy}

\author{Yu-Feng Li}
\email{liyufeng@ihep.ac.cn}
\affiliation{Institute of High Energy Physics, Chinese Academy of Sciences, Beijing 100049, China}
\affiliation{School of Physical Sciences, University of Chinese Academy of Sciences, Beijing 100049, China}

\author{Rong-ping Zhang}
\email{zhangrp@ihep.ac.cn}
\affiliation{Institute of High Energy Physics, Chinese Academy of Sciences, Beijing 100049, China}
\affiliation{School of Physical Sciences, University of Chinese Academy of Sciences, Beijing 100049, China}

\date{\dayofweekname{\day}{\month}{\year} \ddmmyydate\today, \currenttime}

\begin{abstract}
The Reactor Antineutrino Anomaly (RAA)
refers to the deficit observed between the average event rate
measured in reactor antineutrino experiments
with respect to the theoretical prediction.
This anomaly was first identified in 2011 ($2.5\sigma$) as a consequence of the
Huber-Muller reactor antineutrino flux calculation.
It was thought to be resolved in 2021
as a result of previous reactor antineutrino flux calculations,
with a reduction to about $1 \sigma$.
In this work,
we present the RAA
obtained with the latest reactor antineutrino flux calculation
published in 2023 by a French research group,
which was never used before for the calculation of the RAA.
It is the first summation flux model
which includes a comprehensive uncertainty budget.
The result indicates a revival of the
RAA at the level of
$2.2\sigma$.
We also consider the usual simplest explanation of the RAA
by active-sterile neutrino oscillations.
We present the constraints on the oscillation parameters
and we derive a tension of
$3.8
\sigma$
with the results of gallium source experiments
(Gallium Anomaly)
taking into account also the
solar neutrino and KATRIN bounds,
that of the combined short-baseline reactor spectral ratio measurements,
and that of the Daya Bay search for a sub-eV sterile neutrino.
Since the tension may be due to underestimated systematic uncertainties
and the main tension is between the gallium data
and the other data,
we finally present the results of a global analysis
with enlarged gallium uncertainties,
which reduce the global tension to
$1.3
\sigma$.
\end{abstract}

\maketitle

\section{Introduction}
\label{sec:intro}

In this paper we present an updated analysis of the status of the short baseline
Reactor Antineutrino Anomaly (RAA).
We also consider the usual simplest explanation of the RAA
by oscillations due to 3+1 mixing of the electron neutrino $\nu_e$
with a light sterile neutrino $\nu_{s}$
(see the reviews in
Refs.\cite{Gariazzo:2015rra,Giunti:2019aiy,Boser:2019rta,Diaz:2019fwt,Dasgupta:2021ies}).
We discuss the compatibility of the
RAA with
the Gallium Anomaly (GA)~\cite{Cadeddu:2025pue},
the solar bound (SOL)~\cite{Gonzalez-Garcia:2024hmf},
the KATRIN bound~\cite{KATRIN:2025lph},
the combined bound of the
DANSS~\cite{DANSS-ICHEP2022},
NEOS+RENO~\cite{RENO:2020hva},
PROSPECT~\cite{PROSPECT:2024gps},
STEREO~\cite{STEREO:2022nzk}
reactor spectral ratio measurements (SPE)
and that of the
Daya Bay search for a sub-eV sterile neutrino~\cite{DayaBay:2024nip} (DB).
For recent reviews of the RAA and GA see
Refs.\cite{Zhang:2023zif,Elliott:2023cvh}.
We do not consider the other short baseline neutrino anomalies
which involve the muon neutrino:
the LSND anomaly and the MiniBooNE anomaly
(see the reviews in
Refs.\cite{Gariazzo:2015rra,Giunti:2019aiy,Boser:2019rta,Diaz:2019fwt,Dasgupta:2021ies}).

In experiments with commercial reactors
the $\bar\nu_{e}$ flux is produced by the $\beta$ decays
of the fission products of
${}^{235}\text{U}$,
${}^{238}\text{U}$,
${}^{239}\text{Pu}$, and
${}^{241}\text{Pu}$
with the respective typical fractions of about
0.55,
0.09,
0.30, and
0.06.
In experiments with research reactors
the $\bar\nu_{e}$ flux is produced only by the $\beta$ decays
of the fission products of
${}^{235}\text{U}$.
Since the RAA is a deficit of
the number of measured $\bar\nu_{e}$ events with respect to
the calculated number,
the RAA depends mainly
on the calculated ${}^{235}\text{U}$ $\bar\nu_{e}$ flux.

\begin{figure*}
\begin{center}
\includegraphics*[width=\linewidth]{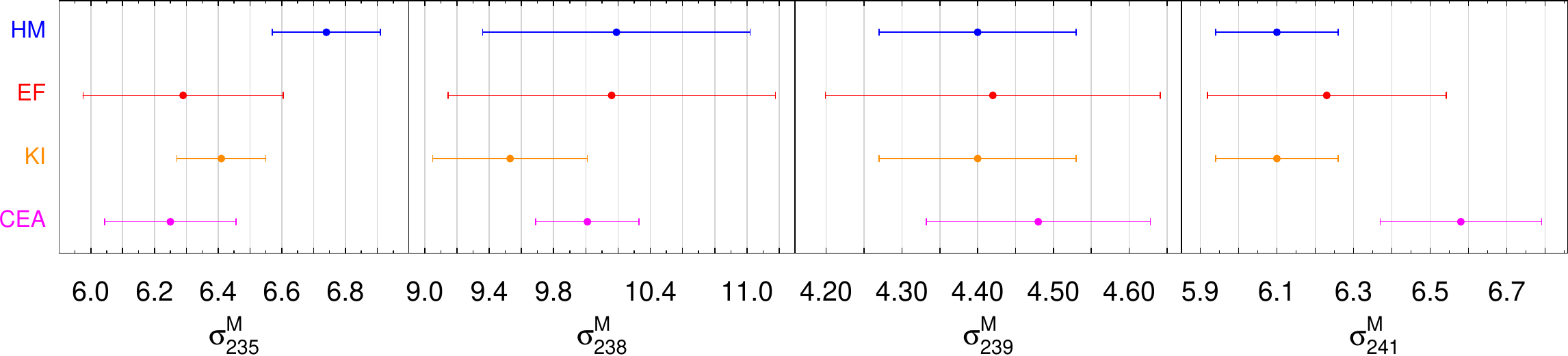}
\end{center}
\vspace{-0.5cm}
\caption{ \label{fig:rea-csf-fig}
Graphical representation of the theoretical IBD yields
in Tab.~\ref{tab:csf} in Appendix~\ref{sub:RAA-csf-the}.
}
\end{figure*}

The reactor antineutrino fluxes
have been calculated with two methods
(see the reviews in Refs.\cite{Huber:2016fkt,Hayes:2016qnu,Zhang:2023zif}):
the summation method which uses the nuclear databases,
and the conversion method in which the neutrino spectra are derived
from the measured electron spectra of the same fission nucleus.

The RAA ($2.5\sigma$ anomaly) was discovered in 2011~\cite{Mention:2011rk}
as a consequence of the calculations of the reactor antineutrino fluxes
in the Huber-Muller (HM) model~\cite{Mueller:2011nm,Huber:2011wv}.
In the HM model the
${}^{235}\text{U}$,
${}^{239}\text{Pu}$, and
${}^{241}\text{Pu}$
fluxes are those
calculated by Huber~\cite{Huber:2011wv} with the conversion method,
using the experimental cumulative $\beta$ spectra
measured at ILL~\cite{VonFeilitzsch:1982jw,Schreckenbach:1985ep}.
The ${}^{238}\text{U}$
flux is that calculated
by Mueller \textit{et al.}~\cite{Mueller:2011nm} with the summation method.
It has been confirmed with the conversion method
using the first and only measurement of the
$\beta$ spectrum of ${}^{238}\text{U}$~\cite{Haag:2013raa}.

In 2019 Estienne, Fallot \textit{et al} (EF)~\cite{Estienne:2019ujo}
presented a new summation method calculation of the reactor antineutrino fluxes
which gave a smaller value for the main
${}^{235}\text{U}$ flux, thus relieving the RAA.

In 2021, Kopeikin, Skorokhvatov, and Titov
of the Kurchatov Institute (KI)~\cite{Kopeikin:2021ugh}
measured the cumulative $\beta$ spectra of
${}^{235}\text{U}$ and ${}^{239}\text{Pu}$
and found that the ratio of the two fluxes is
$ 1.054 \pm 0.002 $
smaller than that of the ILL~\cite{VonFeilitzsch:1982jw,Schreckenbach:1985ep}
measurement on which the HM is based.
Assuming that the HM ${}^{239}\text{Pu}$ antineutrino flux is correct,
the KI group revised the Huber~\cite{Huber:2011wv}
conversion calculation and obtained a smaller
${}^{235}\text{U}$ flux, thus relieving the RAA.

A subsequent analysis~\cite{Giunti:2021kab}
of the experimental reactor antineutrino data
found that the RAA is dramatically
reduced assuming the EF or KI fluxes:
$1.2\sigma$ for EF
and
$1.1\sigma$ for KI.
This suggested a potential resolution of the RAA.

In this article we consider the accurate calculation of the
reactor antineutrino fluxes with the summation method
published in 2023~\cite{Perisse:2023efm},
which was never used before for the calculation of the RAA.
We label this calculation as CEA,
from the affiliations of the authors.
In Section~\ref{sec:RAA}, we discuss the effects of the CEA fluxes on the RAA.

In Section~\ref{sec:3+1} we consider the explanation of the
CEA-RAA
in terms of 3+1 $\nu_{e}$-$\nu_{s}$ oscillations,
where $\nu_{s}$ is a light sterile neutrino
(see the reviews in
Refs.\cite{Gariazzo:2015rra,Giunti:2019aiy,Boser:2019rta,Diaz:2019fwt,Dasgupta:2021ies}).
We emphasize that active-sterile neutrino oscillations
is the only known explanation of the short baseline anomalies which is not ad-hoc.
It is a general possibility which can be tested in many
experiments and was not invented for a specific anomaly.
The existence of sterile neutrinos was proposed many years ago
(see the review in Ref.\cite{Bilenky:1987ty}),
independently of any anomaly.
Hence it is worth to insist on the active-sterile neutrino oscillation
explanation of the short-baseline neutrino data
in spite of the tensions between the existing data in this scenario
\cite{Coloma:2020ajw,Giunti:2021kab,Berryman:2021yan,Giunti:2022btk,Brdar:2023cms,Giunti:2023kyo,Gonzalez-Garcia:2024hmf,Rodrigues:2025tha}.
							
In Section~\ref{sec:3+1}
we discuss the compatibility of the RAA,
the GA~\cite{Cadeddu:2025pue},
the solar bound~\cite{Gonzalez-Garcia:2024hmf},
the KATRIN bound~\cite{KATRIN:2025lph},
the combined bound of the
DANSS~\cite{DANSS-ICHEP2022},
NEOS+RENO~\cite{RENO:2020hva},
PROSPECT~\cite{PROSPECT:2024gps},
STEREO~\cite{STEREO:2022nzk}
reactor spectral ratio measurements,
and that of the
Daya Bay search for a sub-eV sterile neutrino~\cite{DayaBay:2024nip}.
in the framework of 3+1 $\nu_{e}$-$\nu_{s}$ oscillations.
For the GA we consider the average of the (p,n) and
$(^{3}\text{He},^{3}\text{H})$
cross sections in Ref.\cite{Cadeddu:2025pue}.
We discuss the relevant tensions between different data sets
given by the
parameter goodness-of-fit (PG)~\cite{Maltoni:2003cu}.

Assuming the simplest 3+1 $\nu_{e}$-$\nu_{s}$ oscillation explanation
of the RAA and the GA,
in Section~\ref{sec:PG}
we consider the data as measurements of the oscillation parameters.
Inspired by the PDG treatment of incompatible
measures~\cite{ParticleDataGroup:2024cfk},
we calculate the allowed region of the oscillation parameters
with enlarged uncertainties.

\begin{figure}
\begin{center}
\includegraphics*[width=\linewidth]{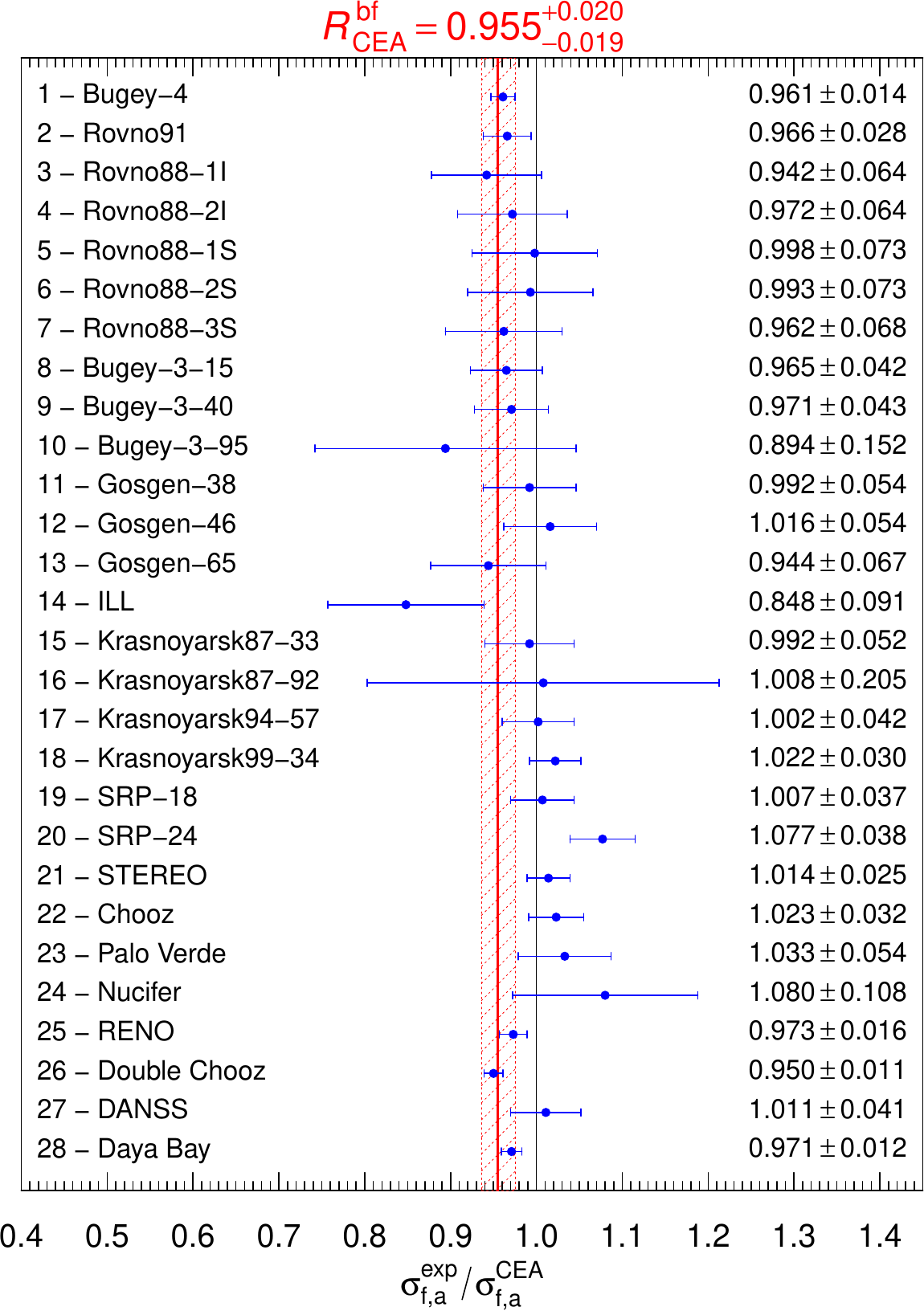}
\end{center}
\caption{ \label{fig:rea-lst-CEA}
Ratios of the measured and predicted yields
of the CEA model for each experiment in Tab.~\ref{tab:rex}
in Appendix~\ref{sub:RAA-csf-exp}.
The uncertainties are only experimental.
The vertical band
and the value given on top show the best-fit ratio $R_{\text{CEA}}^{\text{bf}}$
and its uncertainty
obtained from the minimization of
$\chi^2_{\text{CEA}}$ in Eq.~\ref{eq:chiR}.
}
\end{figure}

\section{Reactor Antineutrino Anomaly}
\label{sec:RAA}

In this Section we update the RAA
considering the CEA~\cite{Perisse:2023efm}
reactor antineutrino fluxes.

The model predictions for the expected experimental results
are expressed by the ``cross section per fission''
$\sigma_{f,a}^{\text{M}}$,
often called ``inverse beta decay (IBD) yield'',
\begin{equation}
\sigma_{f,a}^{\text{M}}
=
\sum_{i}
\sigma_{f,a,i}^{\text{M}}
\quad
\text{with}
\quad
\sigma_{f,a,i}^{\text{M}} = f_{i}^{a} \sigma_{i}^{\text{M}}
,
\label{eq:sfa}
\end{equation}
where $\text{M}$ indicates the model,
$a$ is the experiment number in Tab.~\ref{tab:rex}
in Appendix~\ref{sub:RAA-csf-exp},
and $i$ is the antineutrino flux label:
$i=235$, $238$, $239$, and $241$ for
${}^{235}\text{U}$,
${}^{238}\text{U}$,
${}^{239}\text{Pu}$, and
${}^{241}\text{Pu}$ , respectively.
The cross section per fission
$\sigma_{i}^{\text{M}}$
is the IBD yield for the fission isotope $i$,
and $f_i^a$ is the effective fission fraction of the isotope $i$
for the experiment $a$.
For each fission isotope $i$, the IBD yields are given by
\begin{equation}
\sigma_{i}^{\text{M}}
=
\int_{E_{\nu}^{\text{thr}}}^{E_{\nu}^{\text{max}}}
d E_{\nu}
\,
\Phi_{i}^{\text{M}}(E_{\nu})
\,
\sigma_{\text{IBD}}(E_{\nu})
,
\label{eq:si}
\end{equation}
where $E_{\nu}$ is the neutrino energy,
$\Phi_{i}^{\text{M}}(E_{\nu})$ is the predicted
neutrino flux generated by the fission isotope $i$,
and
$\sigma_{\text{IBD}}(E_{\nu})$
is the detection cross section.
The neutrino energy is integrated from the threshold energy
$ E_{\nu}^{\text{thr}} = 1.806 \, \text{MeV} $
to a maximum value $E_{\nu}^{\text{max}} \ge 8 \, \text{MeV}$.

The IBD yields of the four fission isotopes
predicted by the HM, EF, KI, and CEA models
and their uncertainties
are listed in Tab.~\ref{tab:csf} in Appendix~\ref{sub:RAA-csf-the}
and depicted in
Fig.~\ref{fig:rea-csf-fig}.
As one can see,
the CEA model predicts the smallest value of the main
$\sigma_{235}^{\text{M}}$ yield,
much smaller than the corresponding HM yield.
The value of the $\sigma_{238}^{\text{M}}$ CEA yield
is smaller than those of the HM and EF models
and larger than that of the KI model.
The values of the
$\sigma_{239}^{\text{M}}$ and $\sigma_{241}^{\text{M}}$ CEA yields
are larger than all the others.
Moreover the uncertainties are smaller than those of the EF model
and smaller of those of the HM and KI models only for $\sigma_{238}^{\text{M}}$.

Therefore,
we expect that the CEA-RAA is smaller than the HM RAA.
but it is difficult to estimate if the value of the RAA
is smaller or larger than those of the EF and KI model
without a detailed calculation.

The experimental data that we use are listed in Tab.~\ref{tab:rex}
in Appendix~\ref{sub:RAA-csf-exp},
with the fuel fractions,
the measured cross section per fission,
the relative uncorrelated and correlated experimental uncertainties,
and the source-detector distance.

\begin{table*}
\centering
\begin{minipage}[t]{0.7\linewidth}
\begin{ruledtabular}
\renewcommand{\arraystretch}{1.45}
\begin{tabular}{l|cc|cc|cc}
& \multicolumn{2}{c|}{\bf Rates}
& \multicolumn{2}{c|}{\bf Evolution}
& \multicolumn{2}{c}{\bf Rates+Evolution}
\\
\hline
\bf Model (M)
& $\bm{R_{\text{M}}^{\text{bf}}}$ & \bf RAA
& $\bm{R_{\text{M}}^{\text{bf}}}$ & \bf RAA
& $\bm{R_{\text{M}}^{\text{bf}}}$ & \bf RAA
\\
\hline
\bf HM
&
$0.934 {}^{ + 0.024 }_{ - 0.023 }$
&
$2.6\,\sigma$
&
$0.982 {}^{ + 0.028 }_{ - 0.026 }$
&
$0.7\,\sigma$
&
$0.951 {}^{ + 0.025 }_{ - 0.024 }$
&
$1.9\,\sigma$
\\
\bf EF
&
$0.960 {}^{ + 0.033 }_{ - 0.031 }$
&
$1.2\,\sigma$
&
$0.964 {}^{ + 0.033 }_{ - 0.031 }$
&
$1.1\,\sigma$
&
$0.971 {}^{ + 0.031 }_{ - 0.029 }$
&
$0.9\,\sigma$
\\
\bf KI
&
$0.983 {}^{ + 0.026 }_{ - 0.025 }$
&
$0.6\,\sigma$
&
$0.966 {}^{ + 0.027 }_{ - 0.026 }$
&
$1.2\,\sigma$
&
$0.979 {}^{ + 0.025 }_{ - 0.024 }$
&
$0.9\,\sigma$
\\
\bf CEA
&
$0.955 {}^{ + 0.020 }_{ - 0.019 }$
&
$2.2\,\sigma$
&
$0.957 {}^{ + 0.022 }_{ - 0.021 }$
&
$1.9\,\sigma$
&
$0.955 {}^{ + 0.019 }_{ - 0.018 }$
&
$2.3\,\sigma$
\end{tabular}

\end{ruledtabular}
\end{minipage}
\caption{\label{tab:raa}
Values of the best-fit ratios
of measured and predicted cross sections per fission
and the corresponding RAA
that we obtained for the
HM~\cite{Mueller:2011nm,Mention:2011rk,Huber:2011wv},
EF~\cite{Estienne:2019ujo},
KI~\cite{Kopeikin:2021ugh}, and
CEA~\cite{Perisse:2023efm}
models.
}
\end{table*}

In order to determine the best-fit value of the ratio
$\R$
of measured and predicted IBD yields for each reactor antineutrino flux model M
(M = HM, EF, KI, CEA),
we analyzed the data
with the least-squares function in Eq.(4) of Ref.\cite{Giunti:2021kab}:
\begin{equation}
\chi^2_{\text{M}}
=
\chi^2_{\text{M,exp}}
+
\chi^2_{\text{M,the}}
\label{eq:chiR}
,
\end{equation}
with
\begin{align}
\chi^2_{\text{M,exp}}
=
\null & \null
\sum_{a,b}
\left( \sigma_{f,a}^{\text{exp}} - \R \sigma_{f,a}^{\text{M}} \right)
\left( V^{\text{exp}} \right)^{-1}_{ab}
\nonumber
\\[-0.3cm]
\null & \null
\hspace{1cm}
\times
\left( \sigma_{f,b}^{\text{exp}} - \R \sigma_{f,b}^{\text{M}} \right)
\label{eq:chiRexp}
,
\\
\chi^2_{\text{M,the}}
=
\null & \null
\sum_{i,j}
\left( r_{i} - 1 \right) \left( \widetilde{V}^{\text{M}} \right)^{-1}_{ij} \left( r_{j} - 1 \right)
,
\label{eq:chiRthe}
\end{align}
where
$a,b$ are the experiment labels in Tab.~\ref{tab:rex}
in Appendix~\ref{sub:RAA-csf-exp},
$i,j = 235, 238, 239, 241$,
and
\begin{equation}
\sigma_{f,a}^{\text{M}}
=
\sum_i r_{i} f_i^a \sigma_{i}^{\text{M}}
.
\label{eq:csfth}
\end{equation}
Here $\sigma_{i}^{\text{M}}$
are the four cross sections per fission
of each flux model M in Tab.~\ref{tab:csf} in Appendix~\ref{sub:RAA-csf-the},
and the nuisance parameters $r_{i}$
take into account the respective uncertainties.
The matrix
$\widetilde{V}^{\text{M}}$
is the fractional covariance matrix of the flux model M,
$
\widetilde{V}^{\text{M}}_{ij}
=
V^{\text{M}}_{ij}
/
( \sigma_{i}^{\text{M}} \sigma_{j}^{\text{M}} )
$,
where $V^{\text{M}}$ is the model covariance matrix.
The
${}^{235}\text{U}$,
${}^{239}\text{Pu}$,
${}^{241}\text{Pu}$
elements of the covariance matrix of the HM model have been extracted
from Tables VII-IX of Ref.\cite{Huber:2011wv}.
The ${}^{238}\text{U}$ HM yield is uncorrelated,
since it was calculated separately in Ref.\cite{Mueller:2011nm}.
The EF yields uncertainties are not given in the EF paper~\cite{Estienne:2019ujo}.
Therefore,
as in Ref.\cite{Giunti:2021kab},
we adopt the uncorrelated uncertainties associated with the summation spectra estimated in Ref.~\cite{Hayes:2017res}:
5\% for
$^{235}\text{U}$,
$^{239}\text{Pu}$, and
$^{241}\text{Pu}$,
and 10\% for $^{238}\text{U}$.
We corrected the correlations of the KI model
\label{pag:cor-KI}
which before have been assumed to be equal
to those of the HM model~\cite{Giunti:2021kab}.
As explained in Ref.\cite{Kopeikin:2021ugh},
in the KI model the $^{238}\text{U}$ flux is not that of the HM model,
but that measured in Ref.\cite{Haag:2013raa}
(which is compatible with the HM $^{238}\text{U}$ flux
and has smaller uncertainty),
where the $^{238}\text{U}$ flux is assumed to be fully correlated to the
$^{235}\text{U}$ flux.
Therefore,
since the $^{235}\text{U}$ flux is strongly correlated with the
$^{239}\text{Pu}$ and $^{241}\text{Pu}$ fluxes
as in the HM model,
in the revised KI model all the fluxes are strongly correlated.
The correlations of the CEA yields
are given in the Supplemental Material of Ref.\cite{Perisse:2023efm}.
For further details see the discussion in Appendix~\ref{sec:details}.

The ratio of the measured and predicted yields
of the CEA model for each experiment in Tab.~\ref{tab:rex}
in Appendix~\ref{sub:RAA-csf-exp} are shown in
Figure~\ref{fig:rea-lst-CEA}.
The uncertainties are only experimental.
The vertical band
and the value given on top show the best-fit ratio $R_{\text{CEA}}^{\text{bf}}$
and its uncertainty
obtained from the minimization of the
$\chi^2$ in Eq.~\ref{eq:chiR}.
It takes into account also the
experimental correlations given in Tab.~\ref{tab:rex}
and the covariance of the four cross sections per fission
of the CEA model~\cite{Perisse:2023efm}.

The ``Rates'' columns
in Table~\ref{tab:raa}
show the values of the best-fit ratios
of the HM, EF, KI, and CEA models
and the corresponding RAA.
The value
$2.2\sigma$
of the CEA-RAA
is smaller than the value
$2.6\sigma$
HM RAA.
However, it is almost at the level of the HM RAA\footnote{
The value of the RAA in the original publication~\cite{Mention:2011rk}
which discovered the RAA was $2.5\sigma$.
The small difference with the result of our fit is due to the
reevaluation of the HM yields in Ref.\cite{Giunti:2021kab}
and the update of the experimental data.
}.

The CEA-RAA is larger than those of the EF an KI models.
The reason
is not the correlations of the CEA yields, which are very small and negligible
(see Eq.\eqref{eq:cor-CEA} in Appendix~\ref{sec:details}).
Indeed, without these correlations we obtain the same CEA-RAA of
$2.2\sigma$.

The CEA-RAA is larger than the EF and KI RAA's
because of the covariances of the EF and KI yields.
Indeed, if we replace the CEA covariance with that of the EF model,
we obtain a RAA of
$1.3\sigma$,
which is only slightly bigger than the EF RAA in Tab.~\ref{tab:raa}.
If we replace the practically uncorrelated CEA covariance matrix with
the strongly correlated covariance matrix of the KI model
(see Eq.\eqref{eq:cor-KI} in Appendix~\ref{sec:details}),
we obtain a RAA of
$0.8\sigma$,
which is only slightly bigger than the KI RAA in Tab.~\ref{tab:raa}. 
Thus, a careful validation of the uncertainties and correlations in the predicted isotopic flux yields is essential for accurately resolving the RAA.

We considered also the
Daya Bay evolution data~\cite{DayaBay:2025ngb} only,
with the results in the ``Evolution'' columns of Tab.~\ref{tab:raa}.
With this limit of the data the RAA's decrease,
but it is not negligible for the CEA model
($1.9\sigma$).

As shown in the ``Rates+Evolution'' columns of Tab.~\ref{tab:raa},
with the complete data sets and the Daya Bay evolution data
in replacement to the corresponding total rate measurement,
the HM, EF, and KI RAA's decrease,
whereas the CEA RAA slightly increases to
$2.3\sigma$.
However,
in the following we will not consider the Daya Bay evolution data.

The conclusion of this Section is that
our calculations lead to a revival of the RAA at the level of
$2.2\sigma$.

\begin{figure*}
\centering
\setlength{\tabcolsep}{0pt}
\begin{tabular}{cc}
\subfigure[]{\label{fig:3+1-2s}
\begin{tabular}{c}
\includegraphics*[width=0.49\linewidth]{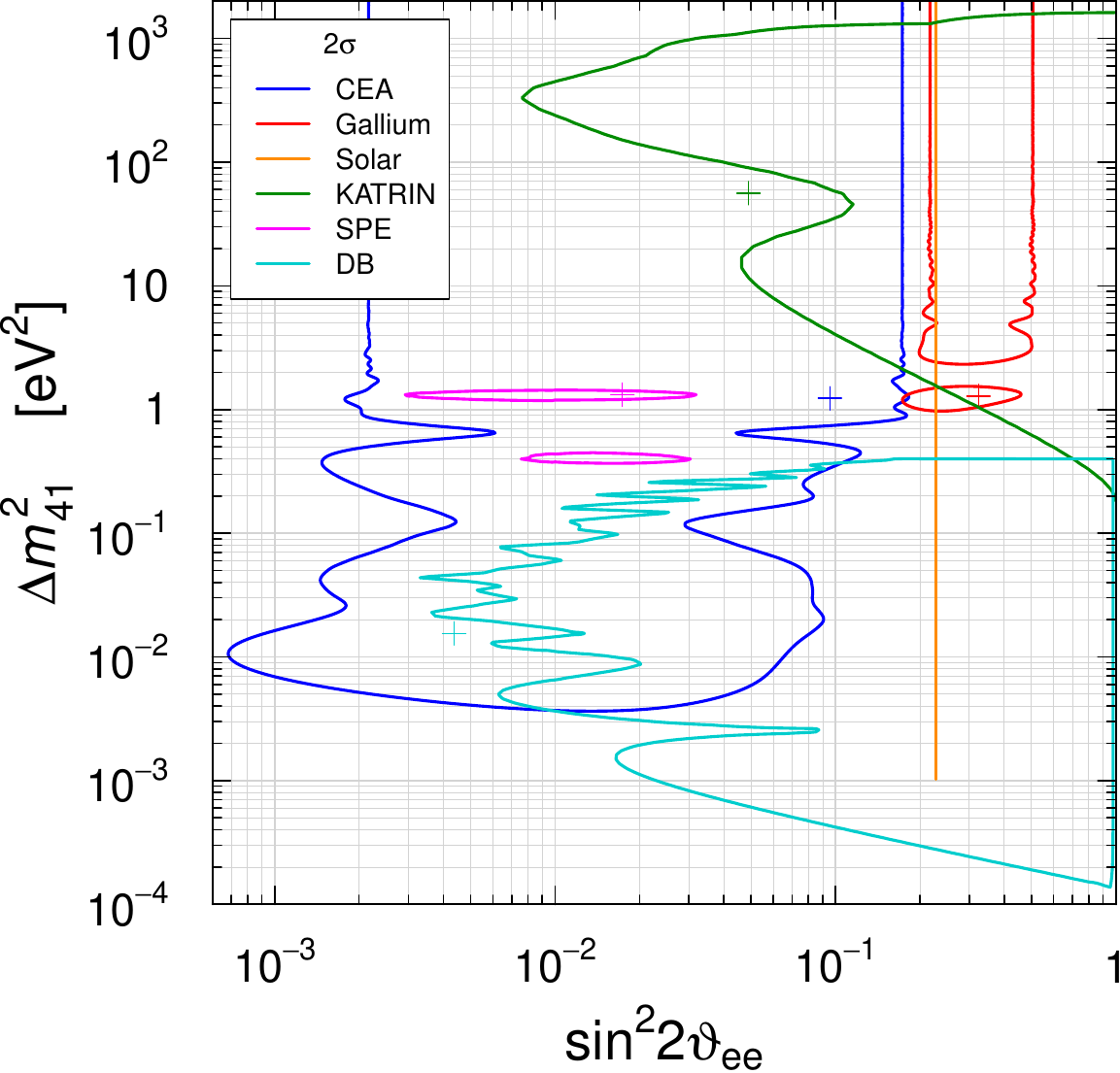}
\\
\end{tabular}
}
&
\subfigure[]{\label{fig:3+1-3s}
\begin{tabular}{c}
\includegraphics*[width=0.49\linewidth]{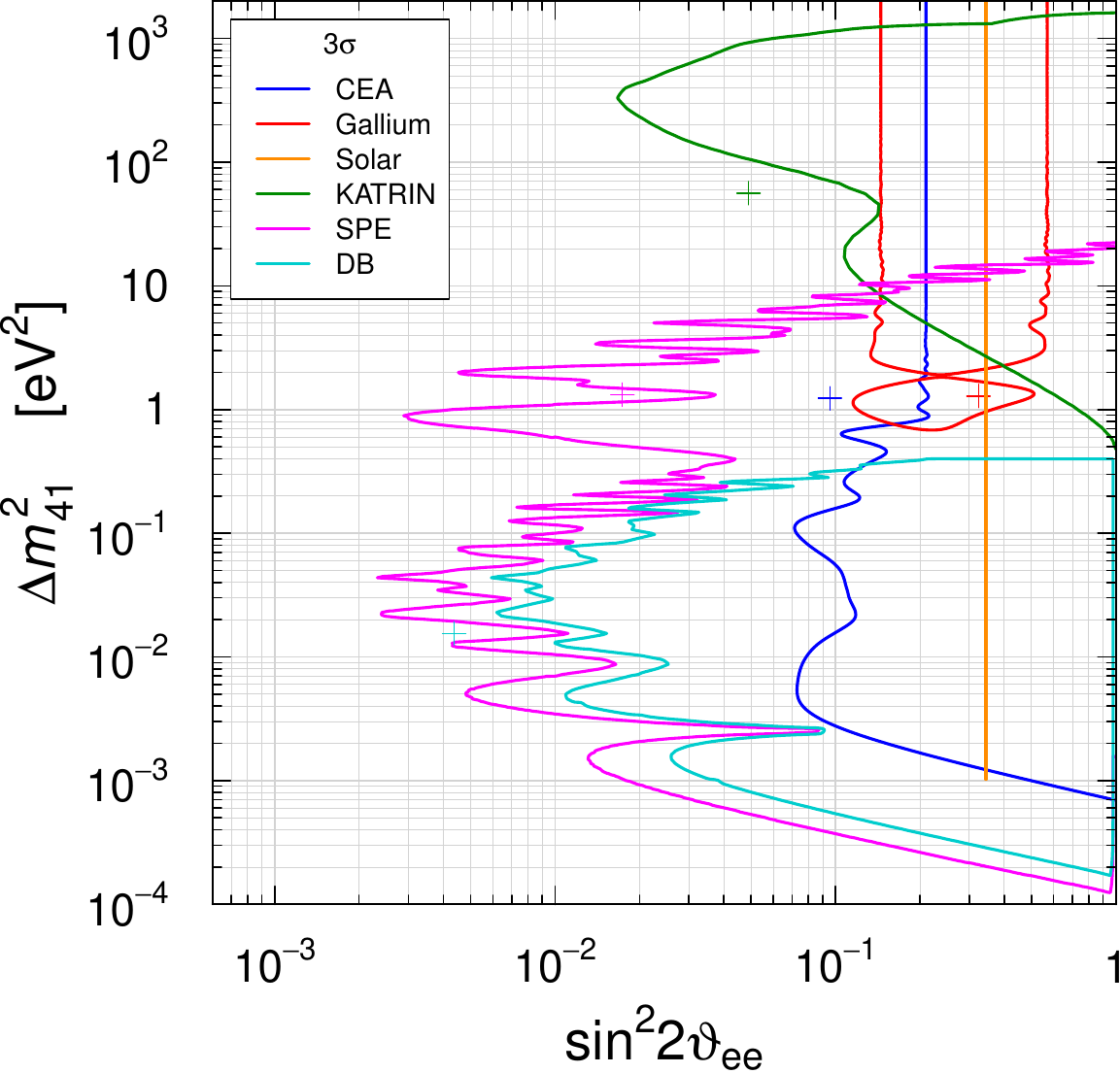}
\\
\end{tabular}
}
\end{tabular}
\caption{\label{fig:3+1}
Contours of the
$2\sigma$~\subref{fig:3+1-2s}
and
$3\sigma$~\subref{fig:3+1-3s}
allowed regions in the
($\sin^2\!2\vartheta_{ee}$,$\Delta{m}^2_{41}$)
plane obtained
from our neutrino oscillation fit of
the reactor antineutrino data in Tab.~\ref{tab:rex}
with the CEA model~\cite{Perisse:2023efm}
and those obtained from
the gallium neutrino data in Ref.\cite{Cadeddu:2025pue},
the solar neutrino data in Ref.\cite{Gonzalez-Garcia:2024hmf},
the KATRIN data~\cite{KATRIN:2025lph},
the combined
DANSS~\cite{DANSS-ICHEP2022},
NEOS+RENO~\cite{RENO:2020hva},
PROSPECT~\cite{PROSPECT:2024gps},
STEREO~\cite{STEREO:2022nzk}
reactor spectral ratio measurements (SPE)
and the
Daya Bay search for a sub-eV sterile neutrino~\cite{DayaBay:2024nip}.
The best fit for each group of experiments is indicated
by a cross with the corresponding color.
}
\end{figure*}

\section{3+1 Oscillations}
\label{sec:3+1}

In this Section we consider the explanation of the
CEA-RAA
in the framework of 3+1 $\nu_{e}$-$\nu_{s}$ oscillations
(see the reviews in
Refs.\cite{Gariazzo:2015rra,Giunti:2019aiy,Boser:2019rta,Diaz:2019fwt,Dasgupta:2021ies}),
which
is the only known explanation of the short baseline anomalies which is not ad-hoc,
as already mentioned in the Introduction.

In the 3+1 framework,
the effective short baseline electron neutrino survival probability
for the experiment $a$ is given by
\begin{equation}
P_{ee}^{a}
=
\left\langle
1 - \sin^2\!2\vartheta_{ee} \, \sin^2 \left(\frac{\Delta m_{41}^2 L}{4E}\right)
\right\rangle
,
\label{eq:Pee}
\end{equation}
where
$\Delta m_{41}^2 = m_{4}^2 - m_{1}^2 \gtrsim 1 \, \text{eV}^2$
is the squared-mass splitting between the non-standard massive neutrino $\nu_{4}$
and the three standard massive neutrinos
$\nu_{1}$,
$\nu_{2}$,
$\nu_{3}$
that have lighter masses with the much smaller
solar and atmospheric
squared-mass splittings
$\Delta m_{21}^2 \approx 7.4 \times 10^{-5} \, \text{eV}^2$
and
$|\Delta m_{31}^2| \approx 2.5 \times 10^{-3} \, \text{eV}^2$
(see Ref.~\cite{ParticleDataGroup:2024cfk}.
The effective mixing angle $\vartheta_{ee}$
is given by
$ \sin^2\!2\vartheta_{ee} = 4 |U_{e4}|^2 ( 1 - |U_{e4}|^2 ) $,
where $U$ of the $4\times4$ mixing matrix.
The average in Eq.\eqref{eq:Pee} is done over the distance range
and the energy spectrum of each experiment $a$
in Tab.~\ref{tab:rex}.

We performed the analysis of the experimental data
with the $\chi^2_{\text{M}}$ in Eq.\eqref{eq:chiR}
by replacing $\R$
with $P_{ee}^{a}$.
The variables of the fit are the 3+1 parameters
$\sin^2\!2\vartheta_{ee}$
and
$\Delta m_{41}^2$
in Eq.\eqref{eq:Pee}.
For each value of these parameters,
$\chi^2_{\text{M}}$
is minimized over the nuisance parameters $r_{i}$.

We obtained the $2\sigma$ and $3\sigma$
allowed regions in the
($\sin^2\!2\vartheta_{ee}$,$\Delta{m}^2_{41}$)
plane
shown in blue in Fig.~\ref{fig:3+1}.
One can see that the $2\sigma$ allowed region has a lower bound for
$\sin^2\!2\vartheta_{ee}$
and the $3\sigma$ allowed region has only an upper bound.
This was expected from the
$2.2\sigma$
CEA-RAA.

In Fig.~\ref{fig:3+1}
we have shown also the regions allowed by the data of the
gallium source experiments in Ref.\cite{Cadeddu:2025pue},
the solar bound~\cite{Gonzalez-Garcia:2024hmf},
the KATRIN bound~\cite{KATRIN:2025lph},
the combined bound of the
DANSS~\cite{DANSS-ICHEP2022},
NEOS+RENO~\cite{RENO:2020hva},
PROSPECT~\cite{PROSPECT:2024gps},
STEREO~\cite{STEREO:2022nzk}
reactor spectral ratio measurements,
and that of the
Daya Bay search for a sub-eV sterile neutrino~\cite{DayaBay:2024nip}.

Remarkably,
at $2\sigma$ the reactor spectral ratio measurements
give small bounded $2\sigma$ allowed regions for
$0.1 \, \text{eV}^2 \lesssim \Delta{m}^2_{41} \lesssim 2 \, \text{eV}^2$.
Similar bounded allowed regions have been found in Ref.\cite{Giunti:2022btk}.
The two small bounded regions at
$\Delta{m}^2_{41} \sim 0.4 \, \text{eV}^2$
and
$\Delta{m}^2_{41} \sim 1.3 \, \text{eV}^2$
are due mainly to the DANSS data.
The two very small bounded regions at
$\Delta{m}^2_{41} \sim 0.11 \, \text{eV}^2$
and
$\Delta{m}^2_{41} \sim 0.15 \, \text{eV}^2$
are due to the combination of the four data sets.

One can also note that for
$\Delta{m}^2_{41} \gtrsim 10^3 \, \text{eV}^2$,
where the KATRIN bound disappears,
there is a bounded $2\sigma$ CEA allowed region
that is excluded only by the gallium data
and by the reactor spectral ratio measurements.
The $2\sigma$ CEA and gallium allowed regions
extend to all values of
$\Delta{m}^2_{41}$ larger than about $10 \, \text{eV}^2$,
because for such large values of
$\Delta{m}^2_{41}$
the oscillation probability is averaged to
$ \overline{P}_{ee} = 1 - 0.5 \sin^2\!2\vartheta_{ee} $.
In this case,
for the CEA reactor analysis,
$ 1 - 0.5 \sin^2\!2\vartheta_{ee} $
corresponds to $R_\text{CEA}$
in the RAA anomaly discussed in Section~\ref{sec:RAA}.
Then, the raster-scan $2\sigma$ allowed interval for
$ \sin^2\!2\vartheta_{ee} $ is $ 0.09 \pm 0.08 $,
which corresponds approximately to the CEA $2\sigma$
allowed interval in Fig.~\ref{fig:3+1-2s}
for large values of $\Delta{m}^2_{41}$.
The equality is approximate because the bounds in Fig.~\ref{fig:3+1-2s}
have not been obtained with the raster-scan method,
but with the usual frequentist
$\Delta\chi^2$ method.

\begin{table}
\centering
\begin{ruledtabular}
\begin{tabular}{lc}
Data Sets & PG
\\
\hline
CEA-RAA+SOL+KATRIN
&
$0.9\sigma$
\\
CEA-RAA+SPE
&
$1.6\sigma$
\\
CEA-RAA+SOL+KATRIN+SPE
&
$1.6\sigma$
\\
CEA-RAA+SOL+KATRIN+SPE+DB
&
$1.4\sigma$
\\
CEA-RAA+GA
&
$2.8\sigma$
\\
GA+SOL+KATRIN
&
$3.0\sigma$
\\
GA+SOL+KATRIN+SPE
&
$3.8\sigma$
\\
CEA-RAA+GA+SOL+KATRIN
&
$3.1\sigma$
\\
CEA-RAA+GA+SOL+KATRIN+SPE
&
$4.0\sigma$
\\
CEA-RAA+GA+SOL+KATRIN+SPE+DB
&
$3.8\sigma$
\\
CEA-RAA+$\overline{\mbox{GA}}$+SOL+KATRIN+SPE+DB
&
$1.3\sigma$
\end{tabular}

\end{ruledtabular}
\caption{\label{tab:pgf}
Values of the parameter goodness of fit (PG)
for different combinations of the data sets
relevant for 3+1 $\nu_{e}$-$\nu_{s}$ oscillations.
We consider the CEA-RAA found in this paper,
the Gallium Anomaly~\cite{Cadeddu:2025pue} (GA),
the solar bound~\cite{Gonzalez-Garcia:2024hmf} (SOL),
the KATRIN bound~\cite{KATRIN:2025lph},
the combined
DANSS~\cite{DANSS-ICHEP2022},
NEOS+RENO~\cite{RENO:2020hva},
PROSPECT~\cite{PROSPECT:2024gps},
STEREO~\cite{STEREO:2022nzk}
reactor spectral ratio measurements (SPE),
and the
Daya Bay search for a sub-eV sterile neutrino~\cite{DayaBay:2024nip} (DB).
The overline on GA in the last row
denotes the enlargement of the gallium uncertainties
discussed in Section~\ref{sec:PG}.
}
\end{table}

To quantify the tensions between different data sets,
we calculated the values
of the parameter goodness of fit (PG)~\cite{Maltoni:2003cu}
for some combinations of the different data sets.
The results are shown in Tab.~\ref{tab:pgf}.
One can see that there is practically no tension
between the CEA-RAA and the solar, KATRIN, and Daya Bay bounds.
There is a moderate tension of
$1.6
\sigma$
between the CEA-RAA and the reactor spectral ratio measurements.
However, the biggest tension is between the GA data and the CEA-RAA, solar and spectral data sets:
from
$2.8
\sigma$
to
$4.0
\sigma$.
Therefore,
the results of the combined fits of the GA data with the other data sets
are not reliable.

\begin{figure}
\begin{center}
\includegraphics*[width=\linewidth]{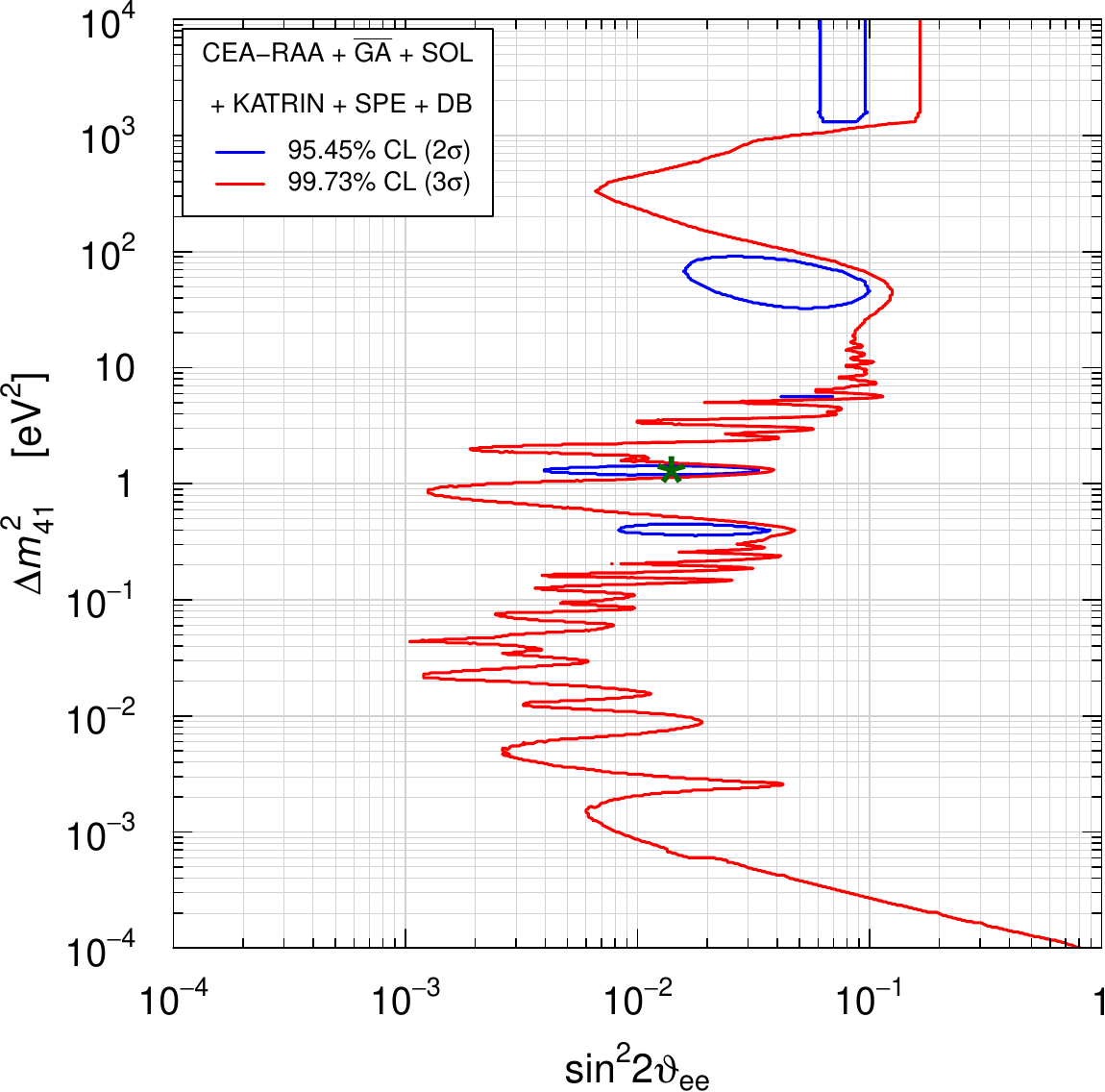}
\end{center}
\caption{ \label{fig:con-tot-2s-3s.pdf}
Contours of the
$2\sigma$ and $3\sigma$
allowed regions in the
($\sin^2\!2\vartheta_{ee}$,$\Delta{m}^2_{41}$)
plane obtained
from the combination of the
reactor (CEA-RAA) analysis
and the bounds of the
solar~\cite{Gonzalez-Garcia:2024hmf} (SOL),
KATRIN~\cite{KATRIN:2025lph},
DANSS~\cite{DANSS-ICHEP2022},
NEOS+RENO~\cite{RENO:2020hva},
PROSPECT~\cite{PROSPECT:2024gps},
STEREO~\cite{STEREO:2022nzk}
reactor spectral ratio (SPE)
experiments,
the
Daya Bay search for a sub-eV sterile neutrino~\cite{DayaBay:2024nip} (DB),
and the gallium~\cite{Cadeddu:2025pue}
analysis with enlarged uncertainties ($\overline{\text{GA}}$)
by the factor $3.8$
given by the PG of the combined fit of all the data sets
(next to last row in Tab.~\ref{tab:pgf}).
The best fit is indicated by a green star.
}
\end{figure}

\section{Global fit with enlarged uncertainties}
\label{sec:PG}

Assuming the simplest 3+1 $\nu_{e}$-$\nu_{s}$ oscillation explanation
of the RAA and the GA,
in this Section
we consider the data as measurements of the oscillation parameters.
Inspired by the PDG treatment of incompatible
measures~\cite{ParticleDataGroup:2024cfk},
we calculate the allowed region of the oscillation parameters
with enlarged uncertainties.

We do not know which set of experiments
can have underestimated the uncertainties.
However, it is plausible that the
reactor, solar, KATRIN, reactor spectral ratio, and Daya Bay data are reliable
and there is no big tension among them.
The main tension is between the gallium data
and
the other data sets.
Therefore,
we consider an increase of the GA uncertainties by the factor
$3.8
\sigma$
given by the PG of the combined fit of all the data sets
(next to last row in Tab.~\ref{tab:pgf}).
This increase of the GA uncertainties
may take into account a possible nuclear physics solution of the GA
with an appropriate transition density~\cite{Cadeddu:2025ueh},
or other explanations~\cite{Arguelles:2022bvt,Hardin:2022muu,Brdar:2023cms,Farzan:2023fqa,Banks:2023qgd}
See, however, the problems of some of these explanations
discussed in \cite{Brdar:2023cms,Giunti:2023kyo}.

The resulting PG is
$1.3
\sigma$
(last line in Tab.~\ref{tab:pgf},
where the GA analysis with enlarged uncertainties
is denoted by $\overline{\mbox{GA}}$).
Since this PG indicates a small global tension,
we performed the combined fit of all the different data sets
(CEA-RAA + $\overline{\mbox{GA}}$ + SOL + KATRIN + SPE + DB)
with the enlarged GA uncertainties.

The results are shown in Fig.~\ref{fig:con-tot-2s-3s.pdf}.
At $3\sigma$ and more there is only an upper bound for
$\sin^2\!2\vartheta_{ee}$,
but at $2\sigma$ there are bounded allowed regions
generated by the CEA-RAA and reactor spectral ratio
$2\sigma$ bounded allowed regions
in Fig.~\ref{fig:3+1-2s}.
The bounded allowed region at $2\sigma$
at $\Delta{m}^2_{41} \sim 10^3 \, \text{eV}^2$
is due to the averaging of the
reactor and gallium oscillation probabilities
for $\Delta{m}^2_{41} \gtrsim 10 \, \text{eV}^2$
and the disappearance of the KATRIN bound
(see the discussion in Section~\ref{sec:3+1}).

We consider the results in Fig.~\ref{fig:con-tot-2s-3s.pdf}
as reliable under the assumption of 3+1 neutrino oscillations
and the reliability of the CEA model.

\section{Summary}
\label{sec:Summary}

In this paper we first examined the current status of the
Reactor Antineutrino Anomaly (RAA)
considering the latest
reactor antineutrino flux calculation published in 2023
(CEA)~\cite{Perisse:2023efm}
and updating the experimental results.
We obtained a
$2.2\sigma$
CEA-RAA, which is larger than those obtained with the
2019 EF~\cite{Estienne:2019ujo}
and
2021 KI~\cite{Kopeikin:2021ugh}
reactor antineutrino flux calculations.
It is almost at the level of the original RAA
discovered in 2011
(HM-RAA~\cite{Mueller:2011nm,Mention:2011rk,Huber:2011wv}).
Therefore, our results indicate a revival of the RAA,
which was reduced to a practically negligible level~\cite{Giunti:2021kab}
after the results of the
EF and KI calculations.

Then we considered the usual explanation of the RAA
in terms of 3+1 $\nu_{e}$-$\nu_{s}$ oscillations,
taking into account also the
Gallium Anomaly (GA)~\cite{Cadeddu:2025pue},
the
solar~\cite{Gonzalez-Garcia:2024hmf} and
KATRIN~\cite{KATRIN:2025lph}
bounds,
the combined bounds of the
DANSS~\cite{DANSS-ICHEP2022},
NEOS+RENO~\cite{RENO:2020hva},
PROSPECT~\cite{PROSPECT:2024gps},
STEREO~\cite{STEREO:2022nzk}
reactor spectral ratio measurements,
and that of the
Daya Bay search for a sub-eV sterile neutrino~\cite{DayaBay:2024nip}.

The CEA allowed region in the
($\sin^2\!2\vartheta_{ee}$,$\Delta{m}^2_{41}$)
plane is compatible with those
obtained 
from the analysis of the solar neutrino data in Ref.\cite{Gonzalez-Garcia:2024hmf},
and from the last results of the KATRIN experiment~\cite{KATRIN:2025lph}.
There is a moderate tension of
$1.6
\sigma$
between the CEA-RAA and the reactor spectral ratio measurements.
However, the biggest tension is between the GA~\cite{Cadeddu:2025pue}
and the CEA-RAA, solar, KATRIN, and reactor spectral ratio data sets:
$2.8
\sigma$
between the GA and the CEA-RAA
and
$3.8
\sigma$
between the GA and the combined
solar, KATRIN and reactor spectral ratio bounds.
The global tension is large:
$3.8
\sigma$.

Therefore, for the calculation of the global fit
we enlarged the GA uncertainties
by the factor $3.8
\sigma$.
We did not enlarge the
CEA-RAA, solar, KATRIN, reactor spectral ratio uncertainties and Daya Bay,
which are more plausible to be reliable
and there is no big tension among them.
For this procedure we have been inspired by the PDG treatment of incompatible
measures~\cite{ParticleDataGroup:2024cfk}.
The resulting global fit has a small global tension of
$1.3
\sigma$.
Therefore,
the results of the analyses of the four different data sets
in the framework of
3+1 $\nu_{e}$-$\nu_{s}$ oscillations
are reconciled.
From this global fit,
we obtained the bounds
shown in Fig.~\ref{fig:con-tot-2s-3s.pdf}.
At $3\sigma$ and more,
we obtained an upper bound for
$\sin^2\!2\vartheta_{ee}$,
whereas at $2\sigma$ there are bounded allowed regions
generated by the CEA-RAA and reactor spectral ratio
$2\sigma$ bounded allowed regions
in Fig.~\ref{fig:3+1-2s}.

In conclusion,
in this paper we found a revival of the RAA
and we calculated reliable allowed regions
in the
($\sin^2\!2\vartheta_{ee}$,$\Delta{m}^2_{41}$)
plane
from the analysis of all the available data
in the framework of 3+1 neutrino oscillations,
assuming the reliability of the CEA model.

\acknowledgements

We thank Thierry Lasserre, Jiajie Ling and Soo-Bong Kim
for useful communications about the KATRIN, Daya Bay and RENO experiments.
We thank also Matteo Cadeddu for useful discussions.
This work was supported in part by the
National Natural Science Foundation of China under grant
number 12075255.

\vspace{0.5cm}
\begin{center}
\bf Appendices
\end{center}
\vspace{-0.5cm}

\appendix

\begin{table*}
\centering
\begin{minipage}[t]{0.7\linewidth}
\begin{ruledtabular}
\begin{tabular}{l|cccc}
\bf
Model (M)
& $\bm{\sigma_{235}^{M}}$
& $\bm{\sigma_{238}^{M}}$
& $\bm{\sigma_{239}^{M}}$
& $\bm{\sigma_{241}^{M}}$
\\
\toprule
\bf HM
&
$6.740 \pm 0.170$
&
$10.190 \pm 0.830$
&
$4.400 \pm 0.130$
&
$6.100 \pm 0.160$
\\
\hline
\bf EF
&
$6.290 \pm 0.315$
&
$10.160 \pm 1.016$
&
$4.420 \pm 0.221$
&
$6.230 \pm 0.312$
\\
\hline
\bf KI
&
$6.410 \pm 0.140$
&
$ 9.530 \pm 0.480$
&
$4.400 \pm 0.130$
&
$6.100 \pm 0.160$
\\
\hline
\bf CEA
&
$6.250 \pm 0.206$
&
$10.010 \pm 0.320$
&
$4.480 \pm 0.148$
&
$6.580 \pm 0.211$
\end{tabular}

\end{ruledtabular}
\end{minipage}
\caption{\label{tab:csf}
The theoretical IBD yields of the four fission isotopes in units of $10^{-43} \text{cm}^{2}/\text{fission}$
predicted by the
HM~\cite{Mueller:2011nm,Mention:2011rk,Huber:2011wv},
EF~\cite{Estienne:2019ujo},
KI~\cite{Kopeikin:2021ugh}, and
CEA~\cite{Perisse:2023efm}
models.
The values of the HM, EF, and KI yields are those reevaluated
in Ref.\cite{Giunti:2021kab}.
The values of the CEA yields are those in the CEA paper~\cite{Perisse:2023efm}.
}
\end{table*}

\section{Reactor Antineutrino Anomaly inputs}
\label{sec:RAA-inputs}

In this appendix we present the two inputs for the calculation of the RAA:
the theoretical cross section per fission in
Subsection~\ref{sub:RAA-csf-the}
and
the measured cross section per fission in
Subsection~\ref{sub:RAA-csf-exp}.

\subsection{Theoretical cross section per fission}
\label{sub:RAA-csf-the}

As explained in the main text,
we consider four theoretical cross section per fission
(IBD yields)
models:
HM~\cite{Mueller:2011nm,Mention:2011rk,Huber:2011wv},
EF~\cite{Estienne:2019ujo},
KI~\cite{Kopeikin:2021ugh}, and
CEA~\cite{Perisse:2023efm}.
They are represented graphically in Fig.~\ref{fig:rea-csf-fig}.
For clarity,
the numerical values are given in Tab.~\ref{tab:csf}.

\subsection{Measured cross section per fission}
\label{sub:RAA-csf-exp}

\begin{table*}
\centering
\resizebox{\textwidth}{!}{
\begin{ruledtabular}
\begin{tabular}{clcccccccc}
$a$
&
Experiment
&
$f^{a}_{235}$
&
$f^{a}_{238}$
&
$f^{a}_{239}$
&
$f^{a}_{241}$
&
$\sigma_{f,a}^{\text{exp}}$
&
$\delta_{a}^{\text{exp}}$ [\%]
&
$\delta_{a}^{\text{cor}}$ [\%]
&
$L_{a}$ [m]
\\
\toprule
$1$	&Bugey-4~\cite{Declais:1994ma}	&$0.538$	&$0.078$	&$0.328$	&$0.056$	&$5.750\pm0.084$	&$1.4$	&\rdelim\}{2}{20pt}[1.4]	&$15$\\
$2$	&Rovno91~\cite{Kuvshinnikov:1990ry}	&$0.614$	&$0.074$	&$0.274$	&$0.038$	&$5.850\pm0.170$	&$2.8$	&                       	&$18$\\
\midrule
$3$	&Rovno88-1I~\cite{Afonin:1988gx}	&$0.607$	&$0.074$	&$0.277$	&$0.042$	&$5.700\pm0.387$	&$6.4$	&\rdelim\}{2}{20pt}[3.1] \rdelim\}{5}{20pt}[2.2]	&$18.0$\\
$4$	&Rovno88-2I~\cite{Afonin:1988gx}	&$0.603$	&$0.076$	&$0.276$	&$0.045$	&$5.890\pm0.387$	&$6.4$	&                                               	&$17.96$\\
$5$	&Rovno88-1S~\cite{Afonin:1988gx}	&$0.606$	&$0.074$	&$0.277$	&$0.043$	&$6.040\pm0.444$	&$7.3$	&\rdelim\}{3}{45pt}[3.1]                        	&$18.15$\\
$6$	&Rovno88-2S~\cite{Afonin:1988gx}	&$0.557$	&$0.076$	&$0.313$	&$0.054$	&$5.960\pm0.440$	&$7.3$	&                                               	&$25.17$\\
$7$	&Rovno88-3S~\cite{Afonin:1988gx}	&$0.606$	&$0.074$	&$0.274$	&$0.046$	&$5.830\pm0.410$	&$6.8$	&                                               	&$18.18$\\
\midrule
$8$	&Bugey-3-15~\cite{Declais:1995su}	&$0.538$	&$0.078$	&$0.328$	&$0.056$	&$5.772\pm0.251$	&$4.2$	&\rdelim\}{3}{20pt}[4.0]                        	&$15$\\
$9$	&Bugey-3-40~\cite{Declais:1995su}	&$0.538$	&$0.078$	&$0.328$	&$0.056$	&$5.809\pm0.257$	&$4.3$	&                                               	&$40$\\
$10$	&Bugey-3-95~\cite{Declais:1995su}	&$0.538$	&$0.078$	&$0.328$	&$0.056$	&$5.345\pm0.909$	&$15.2$	&                                               	&$95$\\
\midrule
$11$	&Gosgen-38~\cite{CALTECH-SIN-TUM:1986xvg}	&$0.619$	&$0.067$	&$0.272$	&$0.042$	&$5.989\pm0.326$	&$5.4$	&\rdelim\}{3}{20pt}[2.0] \rdelim\}{4}{20pt}[3.8]	&$37.9$\\
$12$	&Gosgen-46~\cite{CALTECH-SIN-TUM:1986xvg}	&$0.584$	&$0.068$	&$0.298$	&$0.050$	&$6.090\pm0.324$	&$5.4$	&                                               	&$45.9$\\
$13$	&Gosgen-65~\cite{CALTECH-SIN-TUM:1986xvg}	&$0.543$	&$0.070$	&$0.329$	&$0.058$	&$5.615\pm0.399$	&$6.7$	&                                               	&$64.7$\\
$14$	&ILL~\cite{Kwon:1981ua,Hoummada:1995zz}	&$1.000$	&$0$	&$0$	&$0$	&$5.301\pm0.569$	&$9.1$	&                                               	&$8.76$\\
\midrule
$15$	&Krasnoyarsk87-33~\cite{Vidyakin:1987ue}	&$1.000$	&$0$	&$0$	&$0$	&$6.200\pm0.323$	&$5.2$	&\rdelim\}{2}{20pt}[4.1]	&$32.8$\\
$16$	&Krasnoyarsk87-92~\cite{Vidyakin:1987ue}	&$1.000$	&$0$	&$0$	&$0$	&$6.300\pm1.280$	&$20.5$	&                       	&$92.3$\\
$17$	&Krasnoyarsk94-57~\cite{Vidyakin:1994ut}	&$1.000$	&$0$	&$0$	&$0$	&$6.260\pm0.260$	&$4.2$	&0                      	&$57.3$\\
$18$	&Krasnoyarsk99-34~\cite{Kozlov:1999ct}	&$1.000$	&$0$	&$0$	&$0$	&$6.390\pm0.186$	&$3.0$	&0                      	&$34.0$\\
\midrule
$19$	&SRP-18~\cite{Greenwood:1996pb}	&$1.000$	&$0$	&$0$	&$0$	&$6.293\pm0.231$	&$3.7$	&\rdelim\}{2}{20pt}[2.7]	&$18.2$\\
$20$	&SRP-24~\cite{Greenwood:1996pb}	&$1.000$	&$0$	&$0$	&$0$	&$6.729\pm0.237$	&$3.8$	&                       	&$23.8$\\
\midrule
$21$	&STEREO~\cite{STEREO:2020fvd}	&$1.000$	&$0$	&$0$	&$0$	&$6.340\pm0.159$	&$2.5$	&0	&$9-11$\\
$22$	&Chooz~\cite{CHOOZ:2002qts}	&$0.496$	&$0.087$	&$0.351$	&$0.066$	&$6.117\pm0.191$	&$3.2$	&0	&$\approx 1000$\\
$23$	&Palo Verde~\cite{Boehm:2001ik}	&$0.600$	&$0.070$	&$0.270$	&$0.060$	&$6.253\pm0.327$	&$5.4$	&0	&$\approx 800$\\
$24$	&Nucifer~\cite{NUCIFER:2015hdd}	&$0.926$	&$0.008$	&$0.061$	&$0.005$	&$6.666\pm0.667$	&$10.8$	&0	&$7.2$\\
$25$	&RENO~\cite{RENO:2020dxd}	&$0.571$	&$0.073$	&$0.300$	&$0.056$	&$5.852\pm0.097$	&$1.6$	&0	&$\approx 411$\\
$26$	&Double Chooz~\cite{DoubleChooz:2019qbj}	&$0.520$	&$0.087$	&$0.333$	&$0.060$	&$5.710\pm0.063$	&$1.1$	&0	&$\approx 415$\\
$27$	&DANSS~\cite{Alekseev:2025cib}	&$0.541$	&$0.073$	&$0.332$	&$0.055$	&$6.026\pm0.243$	&$4.1$	&0	&$10.9-12.9$\\
$28$	&Daya Bay~\cite{DayaBay:2025ngb}	&$0.564$	&$0.076$	&$0.304$	&$0.056$	&$5.840\pm0.072$	&$1.2$	&0	&$\approx 578$\\
\end{tabular}

\end{ruledtabular}
}
\caption{ \label{tab:rex}
List of the experiments which measured the absolute reactor antineutrino flux.
For each experiment numbered with the index $a$,
the index $k = 235, 238, 239, 241$
indicate the four isotopes
$^{235}\text{U}$,
$^{238}\text{U}$,
$^{239}\text{Pu}$, and
$^{241}\text{Pu}$,
$f^{a}_{k}$ are the fission fractions,
$\sigma_{f,a}^{\text{exp}}$
is the measured cross section per fission
with the experimental uncertainty,
$\delta_{a}^{\text{exp}}$ is the corresponding relative experimental uncertainty,
$\delta_{a}^{\text{cor}}$ is the relative systematic uncertainty
which is correlated in each group of experiments indicated by the braces,
$\delta_{a}^{\text{the}}$ is the relative theoretical uncertainty
which is correlated among all the experiments,
and
$L_{a}$ is the source-detector distance.
}
\end{table*}

In Fig.~\ref{tab:rex} we present the details of the available experimental measurement that we consider.
They have been obtained by reactor antineutrino experiments
which measured the absolute antineutrino flux at different distances
from the reactor source, which is given in the last column.

We listed first the experiments with multiple measurements with correlations
and then the experiments with a single measurement,
which are obviously uncorrelated.
For convenience the measurements are labeled by the number $a$
which is used in the equations.

The columns 3-6 give the average reactor fuel fractions
of the four fission isotopes
$^{235}\text{U}$,
$^{238}\text{U}$,
$^{239}\text{Pu}$,
$^{241}\text{Pu}$,
for each measurement.
Column 7 gives the measured cross sections per fission
with the experimental uncertainties.

Column 8 gives the total percentual uncertainty
and column 9 gives the percentual part of correlated uncertainty.
Besides the groups of experiments with multiple measurements
which are correlated,
we grouped also the correlated Bugey-4 and the Rovno91 measurements
and the correlated Gosgen and ILL experiments,
obtained respectively with the same detector.

With respect to the experimental data considered in Ref.\cite{Giunti:2021kab},
we updated
the RENO measurement with that in Ref.\cite{RENO:2020dxd}
and
the Daya Bay measurement with that in Ref.\cite{DayaBay:2025ngb},
we added the DANSS measurement in Ref.\cite{Alekseev:2025cib}, and
we took into account the correlation of the two SRP measurements~\cite{Zhang:2013ela}.

\section{Details of the RAA calculation}
\label{sec:details}

\begin{table*}
\centering
$\bf R_{\text{\bf M}}=R_{\text{\bf M}}^{\text{\bf bf}}$
\\
\vspace{0.1cm}
\begin{minipage}[t]{0.6\linewidth}
\begin{ruledtabular}
\begin{tabular}{l|r|r|r|r|c|c}
\bf Model (M)
&
$\bm{r_{235}-1}$
&
$\bm{r_{238}-1}$
&
$\bm{r_{239}-1}$
&
$\bm{r_{241}-1}$
&
$\bm{\chi^2_{\text{\bf M,the}}}$
&
$\bm{\chi^2_{\text{\bf M,exp}}}$
\\
\hline
\bf HM
&
$ 0.0024$
&
$-0.0303$
&
$ 0.0021$
&
$ 0.0024$
&
$0.15$
&
$16.33$
\\
\bf EF
&
$ 0.0421$
&
$-0.0673$
&
$-0.0235$
&
$-0.0051$
&
$1.39$
&
$18.75$
\\
\bf KI
&
$-0.0102$
&
$-0.0236$
&
$-0.0150$
&
$-0.0123$
&
$0.26$
&
$19.59$
\\
\bf CEA
&
$ 0.0430$
&
$-0.0142$
&
$-0.0245$
&
$-0.0058$
&
$2.48$
&
$20.82$
\end{tabular}

\end{ruledtabular}
\end{minipage}
\\
\vspace{0.3cm}
$\bf R_{\text{\bf M}}=1$
\\
\vspace{0.1cm}
\begin{minipage}[t]{0.6\linewidth}
\begin{ruledtabular}
\begin{tabular}{l|r|r|r|r|c|c}
\bf Model (M)
&
$\bm{r_{235}-1}$
&
$\bm{r_{238}-1}$
&
$\bm{r_{239}-1}$
&
$\bm{r_{241}-1}$
&
$\bm{\chi^2_{\text{\bf M,the}}}$
&
$\bm{\chi^2_{\text{\bf M,exp}}}$
\\
\hline
\bf HM
&
$-0.0580$
&
$-0.0976$
&
$-0.0664$
&
$-0.0590$
&
$6.78$
&
$16.49$
\\
\bf EF
&
$-0.0005$
&
$-0.1260$
&
$-0.0491$
&
$-0.0116$
&
$2.61$
&
$18.95$
\\
\bf KI
&
$-0.0233$
&
$-0.0538$
&
$-0.0319$
&
$-0.0276$
&
$1.19$
&
$19.07$
\\
\bf CEA
&
$-0.0097$
&
$-0.0332$
&
$-0.0606$
&
$-0.0148$
&
$4.74$
&
$23.58$
\end{tabular}

\end{ruledtabular}
\end{minipage}
\caption{ \label{tab:dev}
Top:
Values of the deviations from one of the four nuisance parameters $r_{i}$
in Eq.(4) for
$\R=R_{\text{M}}^{\text{bf}}$
and the corresponding values of the two contributions to
$\chi^2_{\text{M}}$ in Eq.(3):
$\chi^2_{\text{M,the}}$ and
$\chi^2_{\text{M,exp}}$.
Bottom:
Same values for
$\R=1$.
}
\end{table*}

\begin{figure*}
\centering
\setlength{\tabcolsep}{0pt}
\begin{tabular}{cc}
\subfigure[]{\label{fig:chi-mod}
\begin{tabular}{c}
\includegraphics*[width=0.49\linewidth]{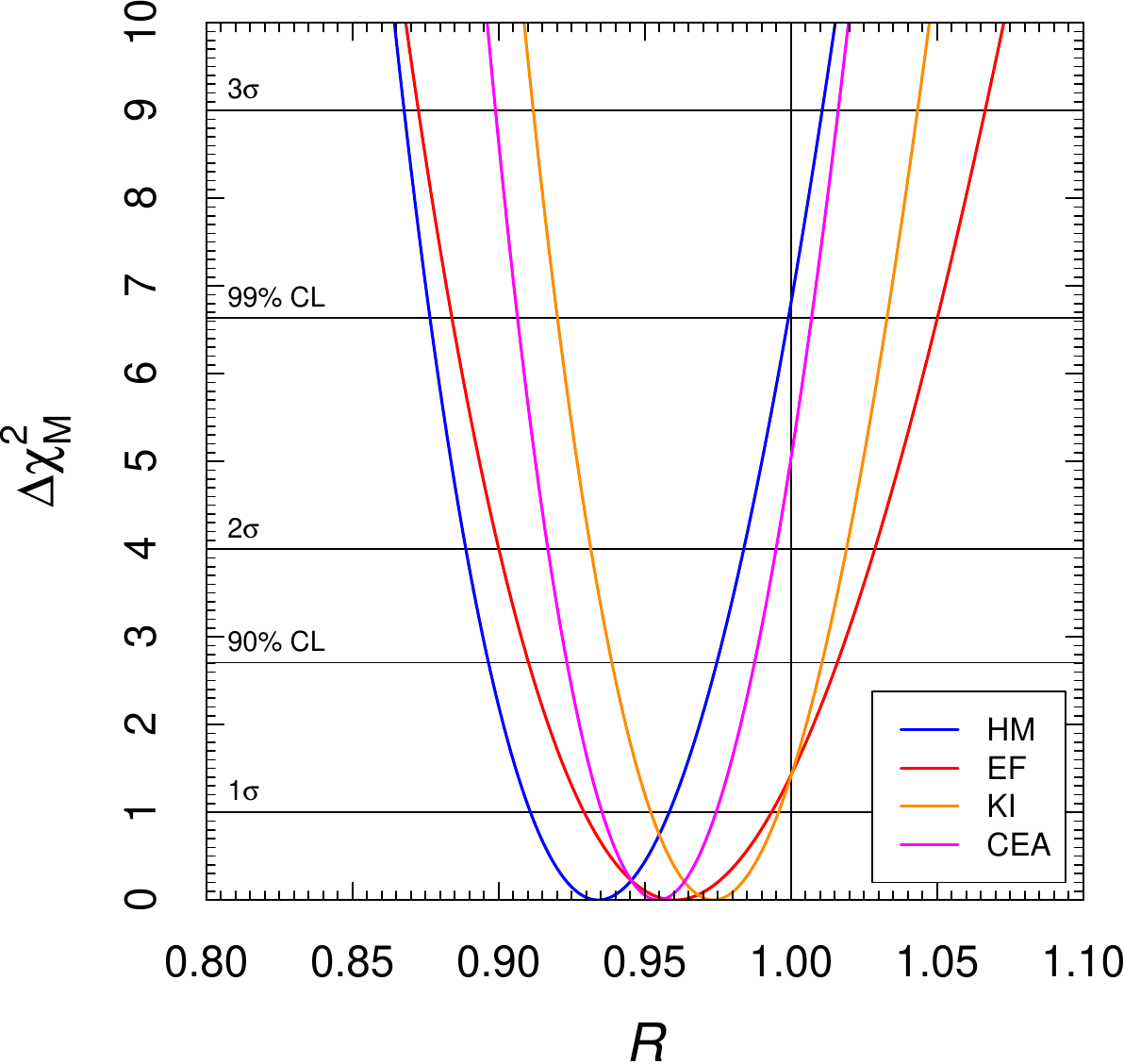}
\\
\end{tabular}
}
&
\subfigure[]{\label{fig:chi-dif}
\begin{tabular}{c}
\includegraphics*[width=0.49\linewidth]{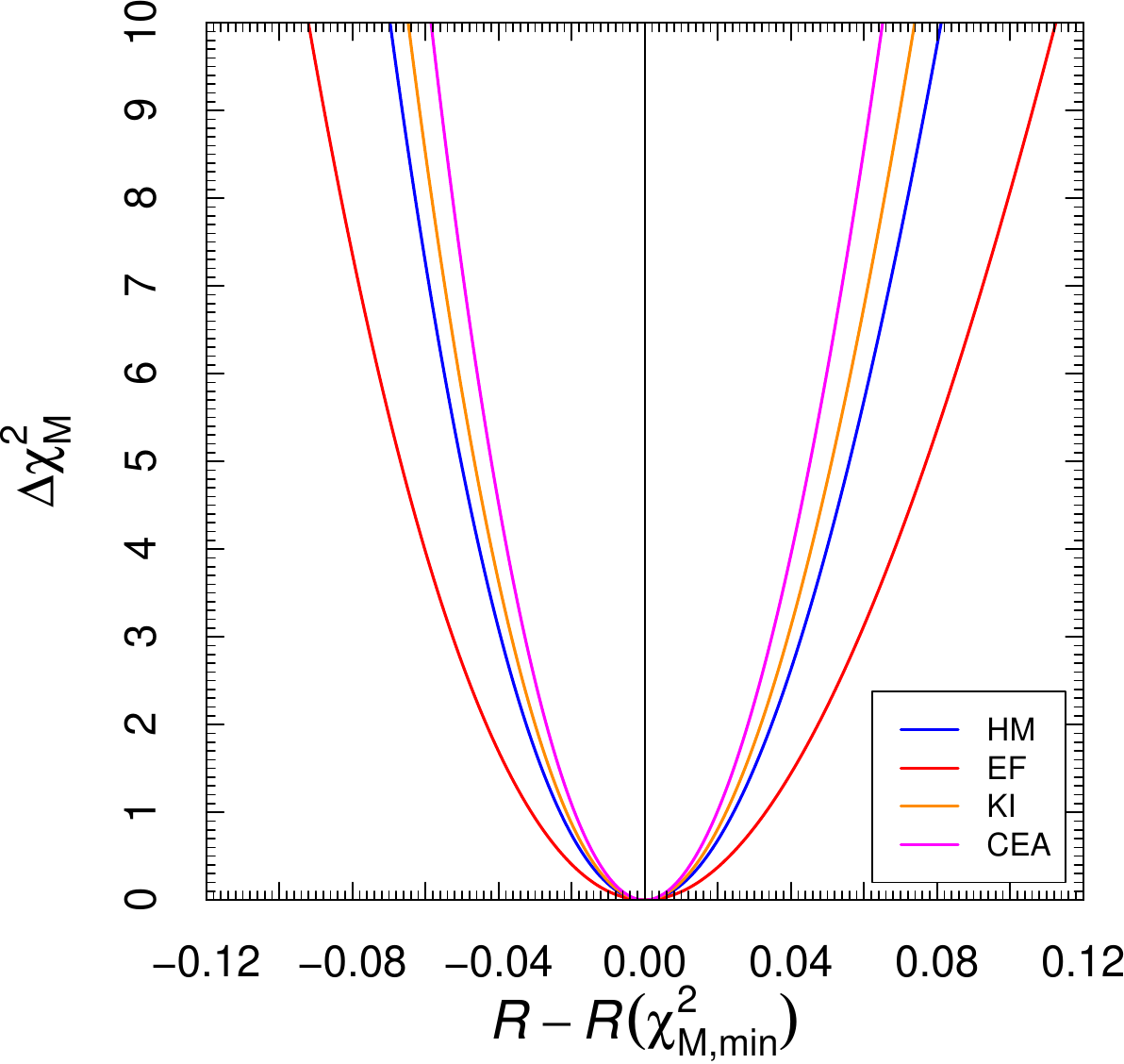}
\\
\end{tabular}
}
\end{tabular}
\caption{\label{fig:chi}
Behaviour of
$\chi^2_{\text{M}}$
for the four flux models
as a function of $R$
\subref{fig:chi-mod}
and of the difference
$R - R(\chi^2_{\text{M,min}})$
\subref{fig:chi-dif},
where $\chi^2_{\text{M,min}}$
is the minimum of $\chi^2_{\text{M}}$ for each model.
}
\end{figure*}

In this Appendix we discuss some details of the calculation of the
"Rates" RAA in Tab.~\ref{tab:raa}.

It is useful to visualize the correlation matrices of the four
reactor flux models that we have considered.
In the HM model there are strong correlations
between the
${}^{235}\text{U}$,
${}^{239}\text{Pu}$, and
${}^{241}\text{Pu}$
fluxes:
\begin{equation}
\begin{pmatrix}
1 & 0 & 0.943 & 0.971  \\
0 & 1 & 0 & 0  \\
0.943 & 0 & 1 & 0.928  \\
0.971 & 0 & 0.928 & 1 

\end{pmatrix}
.
\label{eq:cor-HM}
\end{equation}
In the KI model there are strong correlations
between all the fluxes~\footnote{We corrected
the correlations of the KI model
which before have been assumed to be equal
to those of the HM model~\cite{Giunti:2021kab},
as explained in page~\pageref{pag:cor-KI}.}:
\begin{equation}
\begin{pmatrix}
1 & 0.999 & 0.943 & 0.971  \\
0.999 & 1 & 0.942 & 0.970  \\
0.943 & 0.942 & 1 & 0.928  \\
0.971 & 0.970 & 0.928 & 1 

\end{pmatrix}
.
\label{eq:cor-KI}
\end{equation}
The EF fluxes are totally uncorrelated
and the CEA fluxes have negligible correlations:
\begin{equation}
\begin{pmatrix}
1 & 0.000853 & 0.000906 & 0.000852  \\
0.000853 & 1 & 0.000565 & 0.000531  \\
0.000906 & 0.000565 & 1 & 0.001191  \\
0.000852 & 0.000531 & 0.001191 & 1 

\end{pmatrix}
.
\label{eq:cor-CEA}
\end{equation}
The correlations are important,
because the inverse of a correlated
$\widetilde{V}^{\text{M}}$
in Eq.\eqref{eq:chiRthe}
has negative off-diagonal elements with $i \neq j$
which decrease $\chi^2_{\text{M,the}}$
if $r_{i} - 1$ and $r_{j} - 1$ have the same sign.
This corresponds to the expectation that
if a model has correlated uncertainties
for $r_{i}$ and $r_{j}$,
deviations from one of the same sign
for $r_{i}$ and $r_{j}$
favor the model,
because it confirms the predicted correlations.

The value of the RAA is determined by the $\chi^2$ difference between
$\R=1$ and $\R=R_{\text{M}}^{\text{bf}}$.

Table~\ref{tab:dev} gives the deviations
from one of the four nuisance parameters $r_{i}$
in Eq.(4) for $\R=R_{\text{M}}^{\text{bf}}$ and $\R=1$.

One can see that the absolute values of the deviations are not large,
but they determine a significant increase of
$\chi^2_{\text{M,the}}$ for $\R=1$
with respect to that for $\R=R_{\text{M}}^{\text{bf}}$.
The values of $\chi^2_{\text{M,exp}}$ change, but not much.
The largest increase is
$2.76
$
for the CEA model.
For the KI model there is even a small decrease.

In particular the variations of the absolute values of
the four deviations from $\R=R_{\text{M}}^{\text{bf}}$
to $\R=1$
in the four models are:

\begin{description}

\item[HM]
All positive and significant.
In spite of the large correlations in Eq.\eqref{eq:cor-HM},
we obtain a large increase of $\chi^2_{\text{M,the}}$
and the large RAA in Tab.~\ref{tab:raa}.

\item[EF]
Negative for $r_{235}$
and positive for
$r_{238}$,
$r_{239}$,
$r_{241}$.
Hence, the decrease of
$|r_{235}-1|$ causes a decrease $\chi^2_{\text{M,the}}$,
whereas the increase of
$|r_{238}-1|$,
$|r_{239}-1|$,
$|r_{241}-1|$
cause an increase of $\chi^2_{\text{M,the}}$.
Taking into account also the large uncertainties of the EF yields
(see Fig.~\ref{fig:rea-csf-fig} and Tab.~\ref{tab:csf})
we obtain a small increase of $\chi^2_{\text{M,the}}$
and the negligible RAA in Tab.~\ref{tab:raa}.

\item[KI]
All positive.
Taking into account also the small uncertainties of the KI fluxes
(see Fig.~\ref{fig:rea-csf-fig} and Tab.~\ref{tab:csf})
one would expect a large increase of $\chi^2_{\text{M,the}}$.
Instead,
the strong correlations in Eq.\eqref{eq:cor-KI}
lead to a small increase of $\chi^2_{\text{M,the}}$
and to the negligible RAA in Tab.~\ref{tab:raa}.

\item[CEA]
Negative for $r_{235}$
and positive for
$r_{238}$,
$r_{239}$,
$r_{241}$.
Since the CEA fluxes are practically uncorrelated
(see Eq.\eqref{eq:cor-CEA}),
as the EF fluxes,
but have smaller uncertainties,
the increase of $\chi^2_{\text{M,the}}$ is larger.
It is also larger than that in the KI model
because of the strong correlations of the
$^{235}\text{U}$,
$^{239}\text{Pu}$,
$^{241}\text{Pu}$
KI fluxes.
Taking also into account the significant increase of
$\chi^2_{\text{M,exp}}$,
we obtain the large RAA in Tab.~\ref{tab:raa}.

\end{description}

We finally present in Fig.~\ref{fig:chi-mod}
the dependence from $R$ of
$\chi^2_{\text{M}}$
for the four flux models.
One can see that for each flux model M the amount of the RAA
depends on the value of $R$ for which $\chi^2_{\text{M}}$ is minimum
($\chi^2_{\text{M,min}}$)
and
the slope of $\chi^2_{\text{M}}$.
From low to high values of $R$,
the order of the minima of the four models is
HM, CEA, EF, KI.
In addition,
as shown in Fig.~\ref{fig:chi-dif},
the decreasing steepness order of the four models is
CEA (uncorrelated uncertainties),
KI (smaller but strongly correlated uncertainties of the
$^{235}\text{U}$,
$^{239}\text{Pu}$,
$^{241}\text{Pu}$
IBD yields),
HM (small but strongly correlated uncertainties as KI),
EF (large uncorrelated uncertainties).
The interplay of the distance from one of
$R(\chi^2_{\text{M,min}})$
and the slope of $\chi^2_{\text{M}}$
explain why the values of
$\chi^2_{\text{EF}}$ and $\chi^2_{\text{KI}}$
for $R=1$ are low,
leading to negligible RAA's,
whereas the values of
$\chi^2_{\text{HM}}$ and $\chi^2_{\text{CEA}}$
for $R=1$ are relatively large,
leading to significant RAA's.

\clearpage

\bibliographystyle{apsrev4-1}
\bibliography{revival,tab}

\end{document}